\documentclass[11pt]{article}

\usepackage{amsmath}
\usepackage{amssymb}
\usepackage{amsthm}
\usepackage{latexsym}
\usepackage{color}
\usepackage{graphicx}
\usepackage{appendix}
\usepackage{color}
\usepackage{relsize}
\usepackage{enumerate}
\usepackage[T1]{fontenc}
\usepackage[english]{babel}
\usepackage[table]{xcolor}

\DeclareSymbolFont{calletters}{OMS}{cmsy}{m}{n}
\DeclareSymbolFontAlphabet{\mathcal}{calletters}
\def\be{\begin{eqnarray}}
\def\ee{\end{eqnarray}}

\def\b*{\begin{eqnarray*}}
\def\e*{\end{eqnarray*}}

\newtheorem{Theorem}{Theorem}[part]

\newtheorem{Proposition}{Proposition}[part]

\newtheorem{Corollary}{Corollary}[part]

\makeatletter \@addtoreset{equation}{section}

\@addtoreset{Definition}{section}

\@addtoreset{Theorem}{section}

\@addtoreset{Proposition}{section}

\@addtoreset{Property}{section}

\@addtoreset{Assumption}{section}

\@addtoreset{Corollary}{section}

\@addtoreset{Lemma}{section}

\@addtoreset{Remark}{section}

\@addtoreset{Example}{section}

\@addtoreset{Condition}{section}

\def \R{\mathbb{R}}

\def\={\;=\;}
\def\.{\;.}

\def\1{{\bf 1}}

\def\b*{\begin{eqnarray*}}
\def\e*{\end{eqnarray*}}

 \def\normeL2#1{\left\|{#1}\right\|_{L^2}}

\title{Optimal execution and block trade pricing: a general framework\thanks{This research has been conducted with the support of the Research Initiative ``Exécution optimale et statistiques de la liquidité haute fréquence'' under the aegis of the Europlace Institute of Finance. I would like to thank Yves Achdou (Université Paris-Diderot), Guy Barles (University de Tours), Pierre Cardaliaguet (Université Paris-Dauphine), Vincent Fardeau (Board of Governors of the Federal Reserve System), Nicolas Grandchamp des Raux (HSBC France), Jean-Michel Lasry (University Paris-Dauphine), Anna Obizhaeva (University of Maryland), Jiang Pu (Institut Europlace de Finance), Guillaume Royer (Ecole Polytechnique) and Christopher Ulph (HSBC) for the conversations we had on the subject. I also would like to thank two anonymous referees for their remarks on the paper.}}

\author{Olivier Gu\'eant\footnote{Universit\'e Paris-Diderot, UFR de Math\'ematiques, Laboratoire Jacques-Louis Lions. Avenue de France, 75013 Paris, France. \texttt{gueant@ljll.univ-paris-diderot.fr}}
}

\date{}

\begin{document}

\maketitle

\begin{center}
\textbf{Abstract}
\end{center}

In this article, we develop a general framework to study optimal execution and to price block trades. We prove existence of optimal liquidation strategies and we provide regularity results for optimal strategies under very general hypotheses. We exhibit a Hamiltonian characterization for the optimal strategy that can be used for numerical approximation. We also focus on the important topic of block trade pricing and we propose a methodology to give a price to financial (il)liquidity. In particular, we provide a closed-form formula for the price of a block trade when there is no time constraint to liquidate.

\section*{Introduction}

A general issue for stock traders consists in buying or selling large quantities of shares within a certain time. Unlike for small trades, a trader executing large blocks of shares cannot ignore the significant impact his orders have on the market. A trader willing to sell instantaneously a large quantity of shares would indeed incur very high execution costs or could even be prevented to succeed in selling because of limited available liquidity.\\

In a nutshell, traders face a trade-off between price risk on the one hand and both execution costs and market impact on the other hand. Traders liquidating too fast incur high execution costs but being too slow exposes to possible adverse price fluctuations, effectively leading to liquidation at lower-than-expected prices. For that reason, traders usually split their large orders into smaller ones to be executed progressively over a certain time window.\\

Research on optimal execution\footnote{In this paper, we only focus on liquidation but the issue of buying large quantities of shares can be addressed in the same way.} has long been dedicated to this issue of optimally splitting those large orders. The above trade-off between execution costs, market impact and price risk indeed appeared in the economic and financial literature with Grinold and Kahn \cite{grinold2000active} and has been widely studied since then thanks to the framework developed by Almgren and Chriss in their two seminal papers \cite{almgren1999value,almgren2001optimal}\footnote{Recently, new strands of academic research have developed. Following the seminal paper by Obizhaeva and Wang \cite{obizhaeva2005optimal}, many authors model the dynamics of the order book instead of having a statistical view on execution costs. Also, the focus of research has slightly moved from time scheduling to the actual way to proceed with execution (see \cite{bayraktar2012liquidation,gueant2012general,gueant2012optimal} on limit orders and \cite{kratz2009optimal,kratz2012optimal} on dark pools).}. It is noteworthy that the literature on optimal liquidation started beforehand with Bertsimas and Lo \cite{bertsimas1998optimal}, but their research focused on the minimization of expected execution costs only, consequently ignoring price risk that plays an important part in practice.\\

In the last ten years, the framework proposed by Almgren and Chriss has largely been used in practice and generalized either to better fit real market conditions or to enlarge the scope of modeling possibilities. Initially developed in discrete time with linear execution costs and within a Bachelier model for the price, it has also been considered in continuous time and generalized to allow for nonlinear execution costs and random execution costs \cite{almgren2003optimal}. Black-Scholes dynamics for the price has also been considered and attempts to generalize the model in other directions have also been made, for instance to take account of stochastic volatility and stochastic liquidity \cite{almgren2011optimal}. Discussions on the optimization criterions and their consequences on optimal strategies are also very present in the literature (see for instance  \cite{almgren2007adaptive}, \cite{forsyth2009optimal}, \cite{jk} \cite{lorenz2010mean} and \cite{tse2011comparison}).\\

Here, we consider a Von Neumann-Morgenstern expected utility framework and the specific case of an investor with constant absolute risk aversion.\footnote{Very interesting results in the case of IARA and DARA utility functions are presented in \cite{schied2009risk}. For practical applications, choosing the appropriate level of risk aversion is already a complex task and we shall not go beyond CARA utility functions.} This CARA framework has been studied in \cite{schied2010optimal} in which the author proves that going from deterministic liquidation strategies to stochastic liquidation strategies does not bring any advantage. In this framework, we cover all linear and nonlinear execution cost functions used in practice. In addition to nonlinearity, we cover the case of (deterministic) time-dependent execution costs -- an important case to account for the shape of market volume curves, as the available liquidity is not the same at each hour of the day (for instance in Europe, in addition to the importance of the near-opening and near-closing periods, the opening of the New York marketplace is an important event -- see \cite{ll}). As far as price dynamics is concerned, we focus on the case of a Bachelier model with a drift given by permanent market impact since there is no real difference between Bachelier and Black-Scholes dynamics over short-term horizons. Optimal liquidation strategies are discussed, along with the important problem of block trade pricing. The academic literature indeed focuses on the scheduling of the liquidation process but ignores the global cost of liquidation. We regard the latter issue as an important one, both to know at what price to buy/sell a large portfolio but also for risk management purposes, to evaluate liquidity premia.\\

To give a price to (il)liquidity, we use the indifference pricing approach. In this framework, the price of a block of shares is the minimum price at which an agent would agree to sell shares as a block, to avoid incurring execution costs and price risk. It depends on characteristics of the stock such as volatility and liquidity, but it also depends on the risk aversion of the agent. We provide a closed-form expression for the price of a block trade using results on the long term behavior of viscosity solutions of Hamilton-Jacobi equations.\\

We present the general framework in Section 1. In Section 2, we provide existence, uniqueness and regularity results for the optimal deterministic liquidation strategy. We also provide a Hamiltonian characterization of the optimal strategy and we recall, for the sake of completeness, the result that no stochastic liquidation strategy can improve the best deterministic strategy. Section 3 is dedicated to block trade pricing. We first prove some results about the value function of the control problem. We then determine, in closed-form, the price of a block of shares, given liquidity and market conditions. Section 4 is dedicated to numerical methods to approximate the optimal liquidation strategy. Numerical methods are rarely discussed in the literature (see \cite{forsyth2010hamilton} for one of the rare examples) and we shall see that the Hamiltonian characterization of the optimal strategy provides a good way to approximate the solution. We also present numerical applications for block trade pricing.

\section{Setup and notations}

A trader with a portfolio containing $q_0 > 0$ shares of a given stock is willing to liquidate his portfolio over a time window $[0,T]$.\\

Let us fix a probability space $(\Omega, \mathcal{F}, \mathbb{P})$ equipped with a filtration $(\mathcal{F}_t)_{t\in [0,T]}$ satisfying
the usual conditions. We assume that all stochastic processes are defined on $(\Omega, \mathcal{F},(\mathcal{F}_t)_{t\in [0,T]}, \mathbb{P})$.\\

We introduce the set $\mathcal{P}(t,T)$ of progressively measurable processes defined on $[t,T]$ and the set $\mathcal{A} = \left\lbrace (v_t)_{t\in [0,T]} \in \mathcal{P}(0,T), \int_0^T |v_s| ds \in L^\infty(\Omega) \right\rbrace$ that is the set of admissible execution strategies. For a control process $(v_t)_{t\in [0,T]} \in \mathcal{A}$ -- representing the speed at which the trader sells his shares --, we denote by $(q_t)_{t \in [0,T]}$ the number of shares in the portfolio where $q_t$ is given by:

$$q_t = q_0 - \int_0^t v_s ds.$$

Trades impact the stock market price in two distinct ways. Firstly, there is a permanent market impact (we use the general form of permanent market impact introduced in \cite{gperm}) that imposes a drift to the price process $(S_t)_{t\in [0,T]}$:

$$dS_t = \sigma dW_t - f(|q_0 - q_t|)v_t dt, \qquad \sigma >0,$$ where the function $ f: \mathbb{R}^*_+ \rightarrow \mathbb{R}_+$ is assumed to be nonincreasing and integrable in $0$. In other words, the stock market price decreases when the trader sells his shares ($v_t > 0$) and the impact per share decreases or stays constant as he sells (because $f$ is a nonincreasing function).\footnote{This is in line with the usually observed concavity of permanent market impact -- see \cite{ll}.}\\

Secondly, the price obtained by the trader at time $t$ is not $S_t$ because of execution costs. To model these execution costs (also called instantaneous market impact in the literature), we introduce a strictly convex and even function $L \in C(\mathbb{R}, \mathbb{R}_+)$, increasing on $\mathbb{R}_+$ and verifying the following hypotheses:

$$L(0) = 0, \qquad \lim_{\rho \to + \infty }\frac{L(\rho)}{\rho} = + \infty.$$

These hypotheses are fairly general and, in practice, we want to cover the cases $L(\rho) = \eta |\rho|^{1+\phi}$ for $\eta >0$ and $\phi >0$.\footnote{The initial Almgren-Chriss framework corresponds to $\phi=1$.}\\

We also introduce a market volume process $(V_t)_{t \in [0,T]}$ assumed to be continuous, deterministic and such that $\exists \underline{V} > 0, \overline{V} > 0, \forall t \in [0,T], \underline{V} \le V_t \le \overline{V}$. In this framework, the volume traded by others over the infinitesimal range $[t,t+dt]$ is $V_t dt$. It is assumed to be known in advance, as practitioners use pre-computed volume curves (see \cite{ll}).\\

The execution cost function and the market volume process enter the definition of the cash process $X$ as follows:

$$X_0 = 0, \qquad dX_t = v_t S_t dt - V_t L\left(\frac{v_t}{V_t}\right)dt - \psi |v_t| dt.$$

In financial terms, every time the trader sells $v dt$ shares, he gets the market price of the stock for each share minus execution costs. Execution costs are divided into two parts: a linear part which represents a fixed cost ($\psi \ge 0$) per share -- linked to the bid-ask spread and the tick size --, and a strictly convex and superlinear part modeled by $L$, to account for the cost of trading fast (in proportion of the usual market volume $V_t$). In practice, the higher the participation rate $\frac{v_t}{V_t}$, the higher the executions cost per share.\\

We define our objective function for $v \in \mathcal{A}$ as

$$J(v) = \mathbb{E}\left[-\exp(-\gamma X_T)\right],$$ where $\gamma>0$ is the absolute risk aversion parameter of the trader. This expected utility framework corresponds in the case of Gaussian risks to a classical mean-variance optimization problem. It corresponds also to the optimization criterion of an IS (Implementation Shortfall) in the language of most brokers.\\

The goal of this paper is to solve two problems:

\begin{itemize}
  \item We want to find an optimal strategy to liquidate the portfolio within a certain time $T$. In mathematical terms, we want to maximize $J$ over the set of admissible liquidation strategies $\mathcal{A}_0 = \left\lbrace v \in \mathcal{A}, \int_0^T v_s ds = q_0 \right\rbrace$.
  \item We want to give a price to a block trade. In other words we want a price for a portfolio of $q_0$ shares that takes into account market impact, execution costs and price risk.
  \end{itemize}

\section{Optimal trading curves}

\subsection{Preliminary computations}

In this section, we consider the case of deterministic strategies for optimal liquidation. This means that the trader decides at time $0$ what will be his trading curve for the entire liquidation process.\footnote{In practice, an execution algorithm is usually made of two layers: a strategic layer for the optimal scheduling (trading curve), and a tactical layer, which seeks liquidity inside order books, through all types of orders, and across other (lit or dark) liquidity pools -- see \cite{ll}. Here, we focus on the strategic layer.} Deterministic liquidation strategies are simpler to analyze than general stochastic ones because the cash process at terminal time $T$ (\emph{i.e.} $X_T$) is normally distributed when the liquidation strategy $(v_t)_{t \in [0,T]}$ is deterministic. As we shall see below, it turns out that no stochastic strategy can improve the best deterministic strategy and that the restriction to deterministic strategies is therefore not a real restriction.\\

Let us define the set $\mathcal{A}_{\textrm{det}}$ of deterministic strategies in $\mathcal{A}$ and the set $\mathcal{A}_{0,\textrm{det}}$ of deterministic (liquidation) strategies in $\mathcal{A}_0$. Depending on whether $v$ is in $\mathcal{A}$, $\mathcal{A}_{\textrm{det}}$ or $\mathcal{A}_{0,\textrm{det}}$, we have different expressions and properties for the cash process. The following proposition and its corollary state these properties.

\begin{Proposition}
\label{dist}
Let us consider $v \in \mathcal{A}$ and $t \in [0,T]$. We have
$$X_t + q_t S_t = q_0 S_0  - q_t \int_0^{q_0-q_t} f(|z|) dz - \int_0^{q_0-q_t} F(z) dz$$$$ - \int_0^t V_s L\left(\frac{v_s}{V_s}\right) ds - \int_0^t \psi |v_s| ds + \int_0^t \sigma q_s dW_s,$$where $F(z) = \int_0^z f(|y|)dy.$\\
In particular, if $v \in \mathcal{A}_{\textrm{det}}$ and if $\int_0^t V_s L\left(\frac{v_s}{V_s}\right) ds < +\infty$ then $X_t + q_t S_t$ is normally distributed with mean $q_0 S_0 - q_t \int_0^{q_0-q_t} f(|z|) dz - \int_0^{q_0-q_t} F(z) dz - \int_0^t V_s L\left(\frac{v_s}{V_s}\right) ds - \psi \int_0^t |v_s| ds$ and variance $\sigma^2 \int_0^t q^2_s ds$.
\end{Proposition}

A straightforward corollary of Proposition \ref{dist} is that the cash $X_T$ at the end of the liquidation process is normally distributed in the deterministic case when $q_T = 0$.

\begin{Corollary}[Distribution of $X_T$]
Let us consider $v \in \mathcal{A}_{0,\textrm{det}}$. If $\int_0^T V_s L\left(\frac{v_s}{V_s}\right) ds < +\infty$ then $$X_T \sim \mathcal{N}\left(q_0 S_0 - \int_0^{q_0} F(z) dz - \int_0^T V_sL\left(\frac{v_s}{V_s}\right) - \psi \int_0^T |v_s| ds, \sigma^2 \int_0^T  q^2_s ds\right).$$
\end{Corollary}

This result states that the cash process at the end of the liquidation process can be decomposed in 4 components. The term $q_0 S_0$ represents the initial MtM value of the portfolio since we initially have $q_0$ shares to sell. However, because of market impact, the stock market price decreases as we sell and on average we cannot expect to sell at price $S_0$. This permanent market impact effect is represented by the term $\int_0^{q_0} F(z) dz$. It is important to note that this term does not depend on the liquidation strategy. The third component, namely $\int_0^T V_sL\left(\frac{v_s}{V_s}\right) + \psi \int_0^T |v_s| ds$, represents the impact of execution costs. The fourth component is linked to price risk and states that $X_T$ is a random variable with variance $\sigma^2 \int_0^T  q^2_s ds$. In other words, the faster we sell, the smaller the variance of $X_T$.\\

Using the above corollary and the Laplace transform of a normal distribution, we can find a closed-form expression for the objective function $J$:

\subsection{Existence, uniqueness and characterization of a minimizer}

\begin{Corollary}
Let us consider $v \in \mathcal{A}_{0,\textrm{det}}$. Then
$$J(v) = -\exp\left(-\gamma \left(q_0 S_0 - \int_0^{q_0} F(z) dz - \int_0^T V_sL\left(\frac{v_s}{V_s}\right) ds - \psi \int_0^T |v_s| ds - \frac 12 \gamma \sigma^2 \int_0^T  q^2_s ds \right)\right).$$
\end{Corollary}

We can then define a new objective function for $v \in \mathcal{A}_{0,\textrm{det}}$ by $$I(v) = \int_0^T V_sL\left(\frac{v_s}{V_s}\right) ds + \psi \int_0^T |v_s| ds + \frac 12 \gamma \sigma^2 \int_0^T  q^2_s ds,$$ so that
$$J(v) = -\exp\left(-\gamma \left(q_0 S_0 - \int_0^{q_0} F(z) dz - I(v)\right)\right).$$

Then $$\sup_{v \in \mathcal{A}_{0,\textrm{det}} } J(v) = -\exp\left(-\gamma \left(q_0 S_0 - \int_0^{q_0} F(z) dz - \inf_{v \in \mathcal{A}_{0,\textrm{det}}} I(v)\right)\right),$$ and the maximizers of $J$ correspond to the minimizers of $I$.\\

We therefore study the function $I$ but instead of regarding $I$ as a function of $v$, we regard $I$ as a function of $q$. More precisely, we define

$$\begin{array}{ccccl}
\mathcal{I}_\psi &:& AC_{q_0,0}(0,T) &\to& \mathbb{R}_+\\
& & q & \mapsto & \mathlarger{\int}_0^T \left(V_sL\left(\frac{\dot{q}(s)}{V_s}\right) + \psi |\dot{q}(s)| + \frac 12 \gamma \sigma^2  q^2(s) \right) ds,   \\
\end{array}$$

where $AC_{a_1,a_2}(t_1,t_2)$ is the set of absolutely continuous functions $q$ on $[t_1,t_2]$ satisfying $q(t_1) = a_1$ and $q(t_2) = a_2$.\\

The following theorem states that there exists a unique minimizer of $\mathcal{I}_\psi$ in $ AC_{q_0,0}(0,T)$.

\begin{Theorem}[Existence and uniqueness of a minimizer]
\label{minimizer}
There exists a unique minimizer $q^* \in AC_{q_0,0}(0,T)$ of the function $\mathcal{I}_\psi$. This minimizer is a nonnegative and nonincreasing function. It does not depend on $\psi$.
\end{Theorem}

In addition to providing a unique optimal liquidation strategy, the above theorem states that this strategy consists in selling only. Moreover, it does not depend on the linear part of execution costs that models transaction fees and bid-ask spread. This does not mean that liquidation tactics do not depend on fees or bid-ask spreads but that time scheduling should be the same whatever the value of these linear costs at the strategy scale. For that reason, we now consider $\psi=0$.\\

Let us now come to the characterization of the minimizer $q^*$. Most authors use the Euler-Lagrange equation to compute $q^*$. However, there is no guarantee that $q^*$ is smooth enough to use this methodology. Moreover, since we introduced no differentiability assumption on $L$, we must use another (dual and more general) approach. For that purpose, we introduce $H$, the Legendre transform of $L$: $H(p) = \sup_{\rho \in \mathbb{R}} \rho p - L(\rho)$, and we recall that the strict convexity of $L$ implies that $H$ is a $C^1$ function. The general tools developed in \cite{r1} allow to characterize $q^*$ as follows:

\begin{Proposition}[Hamiltonian equations for $q^*$]
\label{hamilton}
There exists an absolutely continuous function $p^*$ such that $(p^*,q^*)$ solves the following system of equations:

\[
\left \{
\begin{array}{c c c}
    \dot{p}(t) & = & \gamma \sigma^2 q(t) \\
    \dot{q}(t) & = & V_t H'(p(t))\\
\end{array}
\right. \qquad  q(0) = q_0, \quad q(T) = 0.
\]
Moreover, if there exists a couple $(p,q)$ of two absolutely continuous functions satisfying the above equations, then $q=q^*$.
\end{Proposition}

In particular, if we focus on the dual variable $p$, we see that the problem boils down to solving the following non-linear elliptic equation with Neumann boundary conditions:

$$ \ddot{p}(t)  =  \gamma \sigma^2 V_t H'(p(t))$$
$$\dot{p}(0) = \gamma \sigma^2 q_0, \quad \dot{p}(T)= 0.$$

Also, a consequence of the differential characterization of the above Proposition is that $q^* \in C^1([0,T])$. If $H$ is not $C^2$, there is however no reason for $q^*$ to be a $C^2([0,T])$ function. We shall see below that there exists cases where $q^* \notin C^2([0,T])$. This means that one has to be careful when considering the Euler-Lagrange equation. It has also important consequences as far as numerical methods are concerned.\\

Now, since $q^*$ is $C^1$, $v^* = -\dot{q}^* \in \mathcal{A}_{0,\mathrm{det}}$ and we can gather all the results to obtain a theorem about the minimizers of $J$.

\begin{Theorem}[Existence and uniqueness of a maximizer for $J$ in $\mathcal{A}_{0,\mathrm{det}}$]
\label{optimaldet}
There exists a unique maximizer $v^*$ of $J$ in $\mathcal{A}_{0,\mathrm{det}}$.\\
Moreover, $t \mapsto v^*_t$ is a continuous and nonnegative function.
\end{Theorem}

\subsection{Examples and role of the parameters}

To exemplify the above results, we consider two cases for which there is a closed-form solution $q^*$. The first case corresponds to the execution cost function $L(\rho) = \eta \rho^2$ introduced by Almgren and Chriss \cite{almgren2001optimal}. In that case we have the following proposition:

\begin{Proposition}[The Almgren-Chriss case]
Let us consider $L(\rho) = \eta \rho^2$. If $V_t = V$, then $q^*$ is given by
$$q^*(t) = q_0 \frac{\sinh\left(\sqrt{\frac{\gamma \sigma^2 V}{2 \eta}}(T-t)\right)}{\sinh\left(\sqrt{\frac{\gamma \sigma^2 V}{2 \eta}}T\right)},$$
and $$v^*(t)  = q_0 \sqrt{\frac{\gamma \sigma^2 V}{2 \eta}} \frac{\cosh\left(\sqrt{\frac{\gamma \sigma^2 V}{2 \eta}}(T-t)\right)}{\sinh\left(\sqrt{\frac{\gamma \sigma^2 V}{2 \eta}}T\right)}$$
\end{Proposition}

This classical example allows to understand the role of the different parameters:

\begin{itemize}
  \item When the risk aversion parameter $\gamma$ increases, the trader has an incentive to execute faster to reduce price risk. In particular, the liquidation is very fast at the beginning when $\gamma$ is large.
  \item The same reasoning applies with respect to $\sigma$. If volatility increases, the trader wants to trade faster.
  \item $\eta$ is a scale parameter for the execution costs paid by the trader. If $\eta$ increases, the trader liquidate more slowly.
  \item The same liquidity effect holds with respect to market volume. If $V$ decreases, the trader liquidate more slowly.
\end{itemize}

The second case we consider corresponds to $L(\rho) = \eta \rho^{2+\delta}, (\delta>0)$, for $q_0$ small. Although unrealistic for most stocks, this case exemplifies the limited smoothness of $q^*$ since $q^*$ is $C^1$ but not $C^2$.

\begin{Proposition}[Super-quadratic execution costs for small inventories]
Let us consider $L(\rho) = \eta \rho^{2+\delta}$ for $\delta > 0$. Let us assume that $V_t = V$.\\
Then, for $q_0 \le \left(\frac{\delta}{2+\delta}\right)^{\frac{2+\delta}{\delta}} T^{\frac{2+\delta}{\delta}} V^{\frac{1+\delta}{\delta}} \left(\frac{\gamma \sigma^2}{2\eta(1+\delta)}\right)^\frac 1\delta$, $q^*$ is given by
$$q^*(t) = {\left(q_0^{\frac{\delta}{2+\delta}} - \frac{\delta}{2+\delta} V^{\frac{1+\delta}{2+\delta}} \left(\frac{\gamma \sigma^2}{2\eta(1+\delta)}\right)^\frac 1{2+\delta} t\right)_+}^{\frac{2+\delta}{\delta}}.$$
\end{Proposition}

Another interest of this example is that it shows that $q^*(t) = 0$ after a certain time independently of $T$ (if $T$ is large enough). This was not the case in the preceding example.

\subsection*{Appendix to Section 2: Deterministic strategies versus stochastic strategies}

To conclude this section, we recall that no stochastic strategy can improve the best deterministic one. The simple proof we present in the appendix is based on the use of Girsanov Theorem and has initially been proposed in \cite{schied2010optimal}.

\begin{Theorem}[Optimality of deterministic strategies]
\label{optisdet}
$$\sup_{v\in \mathcal{A}_0} \mathbb{E}\left[-\exp\left(-\gamma X_T\right)\right] =\sup_{v\in \mathcal{A}_{0,\mathrm{det}}} \mathbb{E}\left[-\exp\left(-\gamma X_T\right)\right].$$
\end{Theorem}

This theorem states that no liquidation strategy can do better than $v^*$, not only in $\mathcal{A}_{0,\mathrm{det}}$ but more generally in $\mathcal{A}_0$. Hence, we solved our first problem.\\

It is important to notice that this result would not hold in the case of a Black-Scholes dynamics for the price. Furthermore, this result is not true anymore outside of the CARA framework or when the market volume process $V$ is assumed to be stochastic. In practice, it is however very convenient to restrict to deterministic strategies as this permits to separate the strategic layer from the tactical one.

\section{Block trade pricing}

In addition to the optimal liquidation strategy, we are interested in the determination of a price for a block trade. Given the state of the market in terms of liquidity and volatility, we want to determine the maximum price at which an agent would be ready to buy a portfolio containing $q_0$ shares when he is forced to liquidate the portfolio on the market within a certain time $T$. In our expected utility framework, this maximum price $P(T,q_0,S_0)$ is obtained using the indifference pricing approach. It writes

$$P(T,q_0,S_0) = -\frac 1\gamma \log\left( - \sup_{v \in \mathcal{A}_0} J(v)\right) = q_0 S_0 - \int_0^{q_0} F(z) dz - \psi q_0 - \mathcal{I}_0(q^*),$$
where $q^*$ is the optimal trajectory associated to the liquidation of a portfolio with $q_0$ shares on the interval $[0,T]$, as in Section 2.\\

This equation for the price of a block trade has 4 distinct parts. $q_0 S_0$ is the initial MtM value of a portfolio with $q_0$ shares. As we liquidate the portfolio, the market price of the stock decreases and this must be taken into account in the price of the block trade. This is the term $\int_0^{q_0} F(z) dz$ linked to permanent market impact. The third term $\psi q_0$ is linked to the linear part of execution costs. Contrary to the first three ones, the fourth term $\mathcal{I}_0(q^*)$ has been optimized upon to mitigate price risk and at the same time limit execution costs.\\

In practice, one can first compute the optimal liquidation strategy $q^*$ and then compute the price $P(T,q_0,S_0)$ using the above formula. This is a two-step approach. Another approach consists in computing directly the price of a block trade. This one-step approach (or direct approach) is based on the value function of the optimal control problem of Section 2 and relies on the theory of first order Hamilton-Jacobi equations.\\

We introduce the value function of the control problem of Section 2 in the case $\psi=0$:

$$\theta_T(\hat{t},\hat{q}) = \inf_{q \in AC_{\hat{q},0}(\hat{t},T)} \mathlarger{\int}_{\hat{t}}^T \left(V_sL\left(\frac{\dot{q}(s)}{V_s}\right)  + \frac 12 \gamma \sigma^2  q^2(s) \right) ds, \qquad \forall (\hat{t},\hat{q}) \in [0,T)\times\mathbb{R}.$$

%Since we only considered positive inventories, this function is a priori defined on $[0,T)\times\mathbb{R}^*_+$. We rather consider it on $[0,T)\times\mathbb{R}$ -- and we notice that $\forall \hat{t}\in [0,T), \theta_T(\hat{t},0) = 0$ and $\forall (\hat{t},\hat{q}) \in [0,T)\times\mathbb{R}, \theta_T(\hat{t},-\hat{q}) = \theta_T(\hat{t},\hat{q})$, because $L$ is even.\\

Since $\theta_T(0,q_0)$ corresponds to the minimum value of the function $\mathcal{I}_0$ (i.e. $\mathcal{I}_0(q^*)$), we can write $P(T,q_0,S_0)$ as $q_0 S_0 - \int_0^{q_0} F(z) dz - \psi q_0 - \theta_T(0,q_0)$. More generally, we can write the price $P(t,q,S)$ to liquidate $q$ shares on a time window of length $t$ when the price of the stock is $S$ as

$$P(t,q,S) =  q S - \int_0^{q} F(z) dz - \psi q - \theta_T(T-t,q), \forall T \ge t$$

We then see that the study of block trade pricing boils down to the study of the value function $\theta_T$.

\subsection{Basic properties of $\theta_T$}

Since $\theta_T$ plays a central role in our results on block trade pricing, we now study the properties of this function.\\

We first start with basic properties of the value function $\theta_T$ with respect to time:

\begin{Proposition}
\label{mono}
$\forall T > 0, \forall \hat{q} \in \mathbb{R}, \theta_T(\cdot,\hat{q})$ is a nondecreasing function on $[0,T)$.
\end{Proposition}

Another way to see this result is to see the final time as a variable and we obtain the following corollary:

\begin{Corollary}
\label{como}
Consider $t \in \mathbb{R}_+$ and $q \in \mathbb{R}$. Then $T \in (t,+\infty) \mapsto \theta_T(t,q)$ is a nonincreasing function.
\end{Corollary}

In financial terms, it means that the risk-liquidity premium associated to a block trade is a nonincreasing function of the time remaining for liquidation. If a trader has to liquidate a large position and is asked to set a premium upfront for the cost and the risk of the liquidation process, the longer the time window to liquidate, the lower the premium.\\

We now turn to a monotonicity result with respect to $q$.\\

\begin{Proposition}
\label{monoq}
$\forall T > 0, \forall \hat{t} \in [0,T), q \in \mathbb{R}_+ \mapsto \theta_T(\hat{t},q)$ is an nondecreasing function.
\end{Proposition}

In financial terms, the larger the block trade, the higher the risk-liquidity premium associated to this block trade. In fact, this premium is a convex (and in practice superlinear) function of the block trade size, as stated in the following Proposition:

\begin{Proposition}[Convexity of $\theta_T(t,\cdot)$]
\label{convex}
$\forall \hat{t} \in [0,T), \theta_T(\hat{t},\cdot)$ is a convex function.
\end{Proposition}

\subsection{Hamilton-Jacobi equation for $\theta_T$}

In order to study $\theta_T$ and especially its behavior as the time horizon $T$ goes to infinity, we need to write a differential characterization for $\theta_T$. This differential characterization is a Hamilton-Jacobi equation. We thereore first prove that $\theta_T$ is locally Lipschitz on $[0,T)\times\mathbb{R}$ and hence continuous in the following Proposition:

\begin{Proposition}[Local Lipschitz property]
\label{lip}
$\forall Q > 0, \forall \epsilon \in (0,T)$, $\theta_T$ is Lipschitz on $[0,T-\epsilon]\times [-Q,Q]$.
\end{Proposition}

Now, we exhibit the Hamilton-Jacobi equation solved by $\theta_T$:

\begin{Proposition}[Hamilton-Jacobi equation]
\label{visco}
$\theta_T$ is a viscosity solution of the Hamilton-Jacobi equation:
$$-\partial_t \theta_T(t,q) - \frac 12 \gamma \sigma^2 q^2 + V_t H(\partial_q \theta_T(t,q)) = 0, \qquad \mathrm{on\;} [0,T) \times \mathbb{R},$$
$$\theta_T(t,0) = 0, \forall t \in[0,T].$$
\end{Proposition}

The above PDE is satisfied on the domain $[0,T)\times\R$. What happens at $t=T$ is that, due to the liquidation constraint, the value function blows up whenever the constraint is not satisfied. More precisely:

\begin{Proposition}[Singularity of $\theta_T$]
\label{singularity}
$$\lim_{t \to T} \theta_T(t,q) =  \begin{cases} 0, & \mbox{if } q = 0, \\ +\infty, & \mbox{otherwise.} \end{cases}$$
\end{Proposition}

The Hamilton-Jacobi equation allows to derive important results for the price of a block trade. In particular, we shall obtain a closed-form expression for the limit of $\theta_T$ as the horizon time $T$ tends to infinity, and hence for the price of a block trade when there is no time constraint to unwind the portfolio.

\subsection{Main results on block trade pricing}

As discussed above the monotonicity results on $\theta_T$, translate into monotonicity results for the risk-liquidity premium:

\begin{Proposition}
For $q>0$, the risk-liquidity premium $qS - P(T,q,S) = \int_0^{q} F(z) dz + \psi q + \theta_T(0,q)$ is an nondecreasing convex function of $q$ and a nonincreasing function of $T$.
\end{Proposition}

Regarding the price of a block trade, in addition to the above natural monotonicity properties, an important question arises. So far, we only defined the price of a block trade when the agent buying the portfolio was constrained to liquidate it on the market within a certain time $T$. It is in fact natural to focus on the time-unconstrained case. To that purpose, one needs to study the asymptotic behavior of $P(T,q,S)$ as $T\to +\infty$, or equivalently the asymptotic behavior of $T \mapsto \theta_T$. This limit will be obtained in closed-form and it constitutes an upper bound to any block trade price for a finite $T$.

\begin{Theorem}[Asymptotic behavior of $\theta_T$]
\label{asympt}
Let us consider the case of a constant market volume curve $V_t=V$. Then:\\
$$ \forall q \ge 0, \forall t \ge 0, \lim_{T \to +\infty} \theta_T(t,q) = \theta_{\infty}(q) = \int_0^{q} H^{-1}\left(\frac{\gamma \sigma^2}{2V} x^2\right) dx,$$
where $H^{-1}$ is the inverse of the restriction to $\mathbb{R}_+$ of the Legendre transform $H$ of $L$.\\
\end{Theorem}

A straightforward consequence of this result is the following Theorem that gives the price of a block trade in closed-form in the time-unconstrained case:

\begin{Theorem}[Closed-form expression]
\label{closed}
Let us consider the case of a constant market volume curve $V_t=V$.\\
The price of a block trade for a portfolio with $q>0$ shares is:
$$P(q,S) := \lim_{T \to +\infty} P(T,q,S) = q S - \int_0^{q} F(z) dz - \psi q - \int_0^{q} H^{-1}\left(\frac{\gamma \sigma^2}{2V} x^2\right) dx,$$
where $H^{-1}$ is the inverse of the restriction to $\mathbb{R}_+$ of the Legendre transform $H$ of $L$.\\
\end{Theorem}

In financial terms, the price of a block trade is the MtM price of the portfolio $qS$, minus a risk-liquidity premium made of 3 parts. The first part $\int_0^{q} F(z) dz$ is linked to permanent market impact and can be computed in closed form using an antiderivative of $F$. The second part $ \psi q$ is linked to proportional execution costs. The third part, namely $\int_0^{q} H^{-1}\left(\frac{\gamma \sigma^2}{2V} x^2\right) dx$, corresponds to both risk and liquidity. It depends on the stock liquidity through $V$ and $H$ (the Legendre transform of the execution cost function $L$) and on the volatility of the stock $\sigma$ -- the respective importance of the two effects being weighted by $\gamma$.\\

As an illustration, we can compute the price of a block trade when the execution cost function $L$ is $L(\rho) = \eta |\rho|^{1+\phi}$:

\begin{Proposition}[Block trade pricing for $L(\rho) = \eta |\rho|^{1+\phi}$]
\label{sss}
Let us suppose that $V_t=V$ and that $L(\rho) = \eta |\rho|^{1+\phi}$.\\
Then:
$$P(q,S) = q S - \int_0^{q} F(z) dz - \psi q - \frac{\eta^{\frac{1}{1+\phi}}}{\phi^{\frac{\phi}{1+\phi}}} \frac{(1+\phi)^2}{1+3\phi} \left(\frac{\gamma \sigma^2}{2V}\right)^{\frac{\phi}{1+\phi}} q^{\frac{1+3\phi}{1+\phi}}.$$
\end{Proposition}

This allows to understand the role played by the parameters:

\begin{itemize}
  \item The more permanent market impact we have, the higher the premium. This is the role played by $F$.
  \item The higher the proportional fees and the wider the bid-ask spread, the higher the premium. This is the role played by $\psi$.
  \item The more risk adverse a trader is, the higher premium he should quote to compensate for the risk. Therefore, the premium is an increasing function of $\gamma$.
  \item Similarly, the more volatile the market, the higher the premium. In a highly volatile market, the trader quotes a high premium to compensate for price risk.
  \item Due to convexity and superlinearity in liquidation cost, the last term exhibits a convex (increasing) and superlinear behavior with respect to the size $q$ of the block trade. This convexity is measured by $\phi$.
  \item As far as $V$ is concerned, the more liquid a market, the lower the premium.
  \item The higher the execution costs (\emph{i.e.} the higher $\eta$), the higher the premium.
\end{itemize}

\section{Numerical methods and applications}

\subsection{A Newton scheme on the Hamiltonian system}

In this section we discuss the numerical methods that may be used to approximate the optimal liquidation strategy. The goal is to propose a robust method to approximate $q^*$. Most practitioners use the Euler Lagrange characterization for $q^*$ and use a shooting method to solve the problem. This method suffers from several issues. Firstly, it does not work for all execution cost functions and it usually requires $L$ to be $C^2$ -- a condition that is not verified for commonly used execution cost functions. Secondly, this method does not generalize to higher dimension when the goal is to liquidate a multi-asset portfolio.\\

The method we propose can be used in any dimension, although we only present it here in dimension 1. Whereas the Euler Lagrange equation requires $L$ to be $C^2$ -- a condition that is never verified by execution cost functions used in practice --, our method requires $H$ to be $C^2$ -- a condition that is verified when $L(\rho) = \eta |\rho|^{1+\phi}$, with $\phi \in (0,1]$--. The first step consists in discretizing the Hamiltonian system of Proposition \ref{hamilton} using the following scheme:

\[
\left \{
\begin{array}{c c l}
    p_{j+1} & = & p_j + \tau \gamma \sigma^2 q_{j+1} \\
    q_{j+1} & = & q_j  + \tau V_{j+1} {H}'(p_j)\\
\end{array}
\right. \qquad  q_0 = q_0, \quad q_J = 0.
\]

where we considered a subdivision $t_0 = 0 \le\ldots \le  t_j = j\tau \le \ldots \le t_J = T $ of $[0,T]$.\\

The choice of this discretization scheme comes from the discrete counterpart of the continuous model presented in this paper. However, since the boundary conditions are on $q_0$ and $q_J$, we need to use a Newton method to find a solution to this system. As a starting point $\left(p^{0},q^0\right)$ for the algorithm, we can consider:

$$q^{0}_j = \left(1-\frac jJ\right) q_0, \qquad p_j^{0} = \tau \gamma \sigma^2 \sum_{i=1}^j q^0_i.$$

Then, we define recursively $(p^{n+1},q^{n+1})$ from $(p^n, q^{n})$ by:

$$p^{n+1} = p^{n} + \delta p^{n+1}, \quad q^{n+1} = q^{n} + \delta q^{n+1},$$

where $(\delta p^{n+1}, \delta q^{n+1})$ solves the following linear system:

\[
\left \{
\begin{array}{c c l}
    \delta p^{n+1}_{j+1} & = & \delta p^{n+1}_j + \tau \gamma \sigma^2 \delta q^{n+1}_{j+1} \\
    \delta q^{n+1}_{j+1} & = & \delta q^{n+1}_j  + V_{j+1} \tau {H}''(-p^n_j) \delta p^{n+1}_j + \left(q^{n}_j - q^{n}_{j+1} + V_{j+1} \tau {H}'(p^n_j) \right)\\
\end{array}
\right.
\]
$$\delta q^{n+1}_0 = 0, \quad \delta q^{n+1}_J = 0,$$

\subsection{Applications}

As an illustration, this algorithm is used to compute the trading curves associated to IS strategies with the following parameters corresponding (roughly) to the French stock Total SA:
$$S = 40, \quad \sigma = 0.5\quad (20\% \textrm{\; in\; annualized\; terms}), \quad F(q) = 4.5\times10^{-6} sgn(q) |q|^{0.75},$$$$ L(\rho) = 0.02 |\rho|^{0.65}, \quad \psi = 0.004, \quad V = 5000000.$$

We first consider the case where $q_0 = 500000$, and $T=1$ day. In other words, the trader has to liquidate $10\%$ of the usual daily volume, over one trading day. We consider this case for 3 different values of the risk aversion parameter\footnote{We could alternatively have considered 3 scenarios for the value of the volatility parameter $\sigma$ or of the value of the usual daily volume $V$. $\gamma$, $\sigma$ and $V$ indeed only appear in the problem through the ratio $\frac{\gamma \sigma^2}{V}$ that measures the relative importance of price risk.} $\gamma$: $\gamma = 10^{-6}$, $\gamma = 5\times10^{-7}$ and $\gamma = 2\times10^{-6}$.\\

The optimal trading curves are the following:

\begin{figure}[htbp]
\center

  % Requires \usepackage{graphicx}
  \includegraphics[width=11cm]{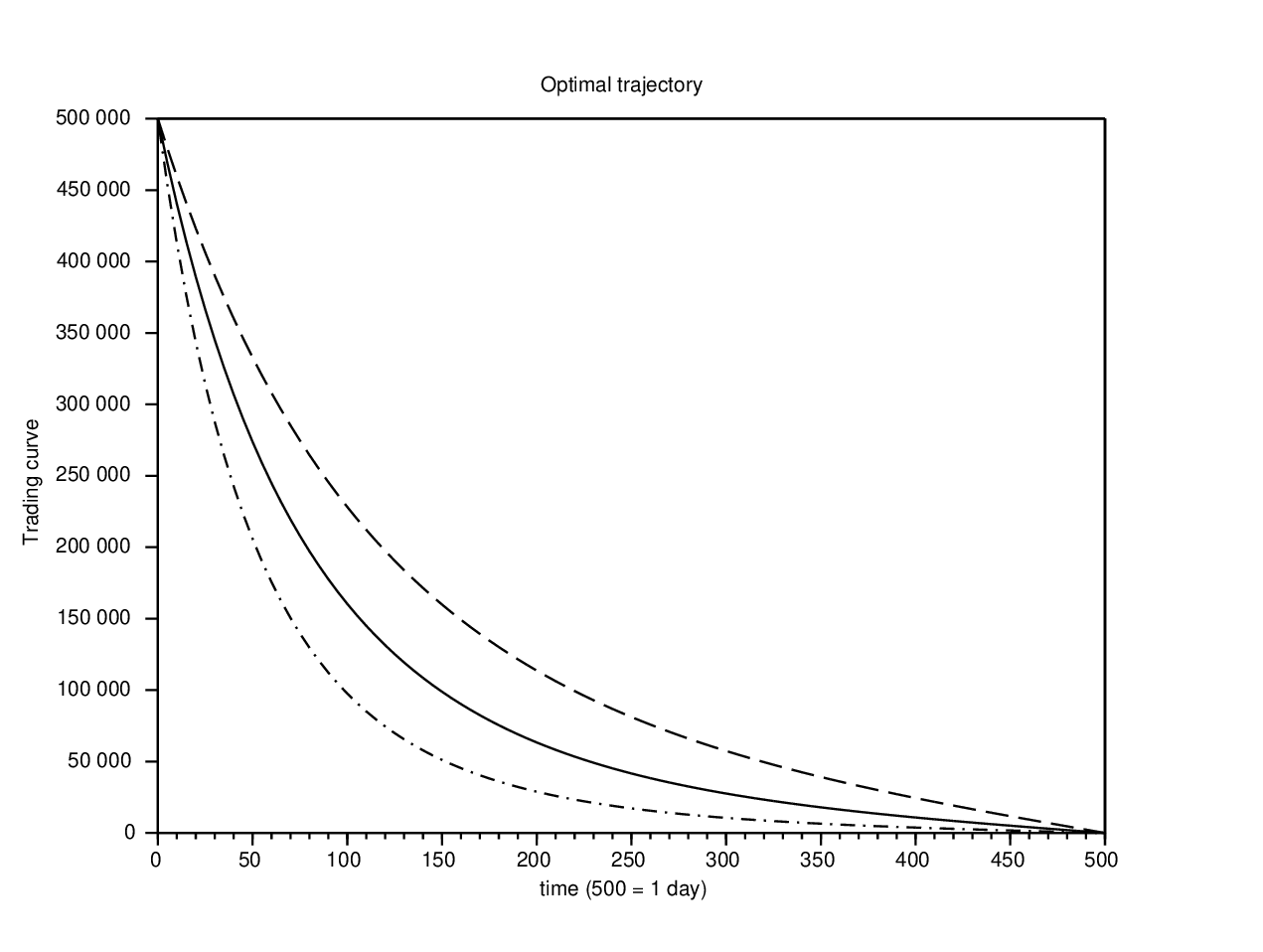}
  \caption{Trading curves for $q_0 = 500000$. The plain line corresponds to $\gamma = 10^{-6}$, the dashed line corresponds to $\gamma = 5\times10^{-7}$, the dash-dotted line corresponds to $\gamma = 2\times10^{-6}$.}
\end{figure}
\vspace{0.8cm}

We see, as expected, that trading curves are convex: the speed at which the trader unwinds his portfolio is high at the beginning and low close to the final time $T$. The trader indeed pays execution costs to reduce price risk at the beginning. Then, as price risk is lower, the trader liquidates more slowly to avoid incurring too much execution cost. In line with this analysis, we also observe that the higher the risk aversion parameter, the faster the liquidation.\\

In order to analyze the associated block trade price, we decompose the risk-liquidity premium into 3 parts:

$$ \underbrace{\int_0^{q} F(z) dz}_{\textrm{permanent\; market\; impact\; (PMI) }}$$
$$+ \underbrace{\psi q}_{\textrm{linear\; execution\; costs \; (LEC)}}$$
$$+ \underbrace{\frac{\eta^{\frac{1}{1+\phi}}}{\phi^{\frac{\phi}{1+\phi}}} \frac{(1+\phi)^2}{1+3\phi} \left(\frac{\gamma \sigma^2}{2V}\right)^{\frac{\phi}{1+\phi}} q^{\frac{1+3\phi}{1+\phi}}}_{\textrm{nonlinear\; execution\; costs\; and\; price\; risk\; (NECPR)}}.$$

Moreover, in addition to this asymptotic value corresponding to $T\to \infty$, we provide the value of the third component computed numerically in our case where $T$ corresponds to one trading day.\\

\begin{tabular}{|c|c|c|c|c|}
\hline
   & PMI & LEC & NECPR ($T \to \infty$) & NECPR ($T=1$) \\
   \hline
  low $\gamma$ ($5\times10^{-7}$)  & $24175 \textrm{\;i.e.\;} 12$ bp & $2000 \textrm{\;i.e.\;} 1$ bp & $5263 \textrm{\;i.e.\;} 2.6$ bp & $5375 \textrm{\;i.e.\;} 2.7$ bp \\\hline
  normal $\gamma$ ($10^{-6}$) & $24175 \textrm{\;i.e.\;} 12$ bp & $2000 \textrm{\;i.e.\;} 1$ bp & $6915 \textrm{\;i.e.\;} 3.5$ bp & $7081 \textrm{\;i.e.\;} 3.5$ bp \\\hline
  high $\gamma$ ($2\times10^{-6}$) & $24175 \textrm{\;i.e.\;} 12$ bp & $2000 \textrm{\;i.e.\;} 1$ bp & $9087 \textrm{\;i.e.\;} 4.5$ bp & $9408 \textrm{\;i.e.\;} 4.7$
   bp\\\hline
\end{tabular}
\vspace{0.8cm}

In the intermediate case where $\gamma = 10^{-6}$, we see that the trader would accept to buy a block of $q_0$ shares if a discount of at least $16.5 = 12 + 1 + 3.5 $ basis points is applied to the MtM price. We also see that the choice of the risk aversion parameter is important and that a conservative choice corresponds to a high $\gamma$. Interestingly, we can also consider our numerical applications as a way to choose $\gamma$. It is indeed straightforward to implicit $\gamma$ from a discount thought to be the minimum necessary one by a trader.\\

Another important fact is that the closed form formula corresponding to $T \to \infty$ is a good approximation for our horizon of $1$ day. This is because one day is usually enough to unwind a position corresponding to $10\%$ of the daily volume.\\

\begin{figure}[!h]
\label{f2}
\center
  % Requires \usepackage{graphicx}
  \includegraphics[width=11cm]{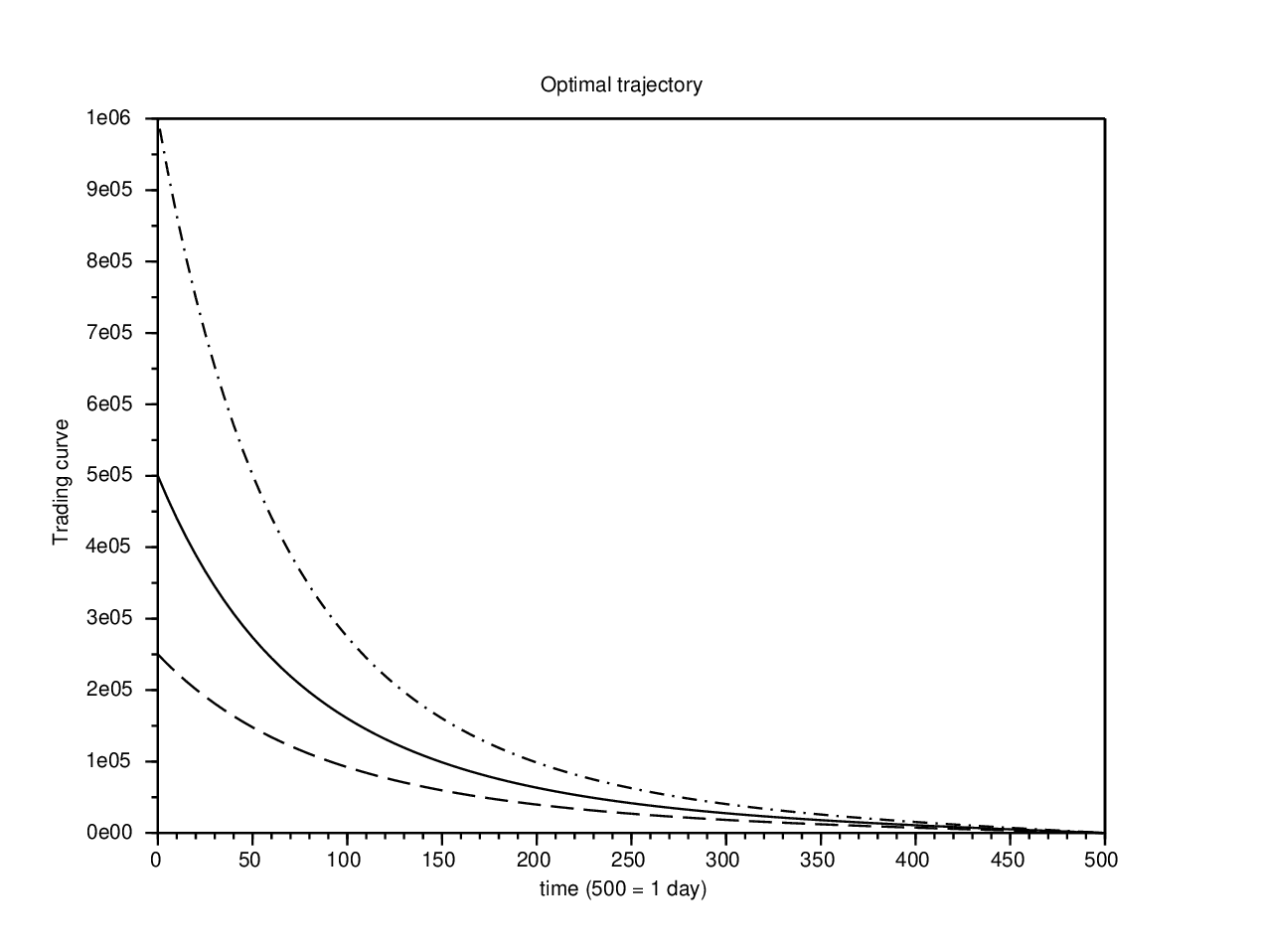}
  \caption{Trading curves for $\gamma = 10^{-6}$. The plain line corresponds to $q_0 = 500000$, the dashed line corresponds to $q_0 = 250000$, the dash-dotted line corresponds to $q_0 = 1000000$.}
\end{figure}

To continue the analysis, we consider different sizes of portfolios to be unwound over 1 trading day. $q_0$ is now $5\%$, $10\%$ or $20\%$ of the usual daily volume. Using the algorithm described above, we obtain the trading curves plotted on Figure \ref{f2}.\\

As expected, the initial speed at which liquidation occurs is an increasing function of $q_0$. This is the consequence of price risk.\\

Coming to risk-liquidity premia, we obtain the following results ($\gamma = 10^{-6}$):\\

\begin{tabular}{|c|c|c|c|c|}
\hline
   & PMI & LEC & NECPR ($T \to \infty$) & NECPR ($T=1$) \\
   \hline
  $q_0 = 5\% \times V$   & $ 7187\textrm{\;i.e.\;} 7.2$ bp & $1000 \textrm{\;i.e.\;} 1$ bp & $ 2003\textrm{\;i.e.\;} 2$ bp & $ 2046 \textrm{\;i.e.\;} 2$ bp \\\hline
  $q_0 = 10\% \times V$  & $24175 \textrm{\;i.e.\;} 12$ bp & $2000 \textrm{\;i.e.\;} 1$ bp & $ 6915 \textrm{\;i.e.\;} 3.5$ bp & $ 7081 \textrm{\;i.e.\;} 3.5$ bp \\\hline
  $q_0 = 20\% \times V$  & $ 81316\textrm{\;i.e.\;} 20.3$ bp & $4000 \textrm{\;i.e.\;} 1$ bp & $ 23881\textrm{\;i.e.\;} 6$ bp & $ 24528 \textrm{\;i.e.\;} 6.1$
   bp\\\hline
\end{tabular}
$$$$
We see that the risk-liquidity premium (in absolute value) is not only increasing with $q_0$ but convex and superlinear in $q_0$. The risk-liquidity premium indeed increases with $q_0$ in basis point. For instance, the total discount a trader would require to buy shares as a block goes from 16.5bps for $10\%$ of the usual daily volume to more than 27 bp for $20\%$ of the usual daily volume.\\

\section*{Conclusion and possible extensions}

This article was motivated by the lack of a general framework for liquidation problems and more importantly by the need to give a price to block trades. We developed such a general framework and proved existence and regularity results for optimal liquidation strategies. Regarding block trade pricing, we provided completely new results that now permit to quantify risk-liquidity premia associated to the liquidation of a portfolio. Specifically, we provided a closed-form formula for the price of a block trade when there is no time constraint to liquidate. We also complemented our results with important remarks concerning the numerical approximations of optimal liquidation strategies and block trade prices.\\

The results we obtained can be generalized straightforwardly when one introduces a constant drift in the price process to model the view of the trader on the evolution of the stock price. Generalizing the results to the case of a multi-asset portfolio is also simple as far as the optimal trajectory is concerned. However, there is no hope to find a closed-form expression for the price of a block trade in the multi-asset case when the correlation matrix is not diagonal. In the multi-asset case, risk-liquidity premia can be computed numerically using the two-step approach that consists in computing first the optimal strategy $q^*$ and then the value of the objective function at $q^*$. For that purpose, a generalization of the Hamiltonian system of Proposition \ref{hamilton} can easily be derived. Another simple extension consists in introducing maximum participation constraints for liquidation. The dual approach used to derive the Hamiltnoian system of Proposition \ref{hamilton} is in that case absolutely necessary, as the Euler-Lagrange equation is degenerated.\\

We believe that the results we have obtained on risk-liquidity premia are important outside of optimal execution and brokerage. In portfolio management, risk-liquidity premia should indeed be used to penalize illiquid portfolios. The framework we developed may also be used to incorporate liquidity features into option pricing, especially to price corporate deals with large nominals.

\section*{Appendix A}

In this appendix, we continue the study of the value function $\theta_T$ and we show under very mild assumptions that $\theta_T$ is a $C^1$ function (and therefore that $\theta_T$ is a classical solution of the Hamilton-Jacobi equation of Proposition \ref{visco}.\\

A first step consists of proving that the value function is semi-concave in the variable $q$.\\

\textbf{Proposition} (Local semi-concavity of $\theta_T(t,\cdot)$)
Let us assume that $L$ is locally semi-concave.\footnote{This is true as soon as $L$ is $C^1$ and verified for all commonly used execution cost functions.}\\
Then, $\forall \hat{t} \in [0,T), \theta_T(\hat{t},\cdot)$ is a locally semi-concave function.\\

\emph{Proof:}\\

Let us consider $Q >0$, $\hat{q}_1 < \hat{q}_2 \in [-Q,Q]$ and $\lambda \in (0,1)$.\\

Let us define $q_1 = q^*_{ \hat{t}, \lambda \hat{q}_1 + (1-\lambda)\hat{q}_2 } - (1-\lambda)\frac{T-t}{T-\hat{t}}(\hat{q}_2-\hat{q}_1)$ and $q_2 = q^*_{ \hat{t}, \lambda \hat{q}_1 + (1-\lambda)\hat{q}_2 } + \lambda\frac{T-t}{T-\hat{t}}(\hat{q}_2-\hat{q}_1)$. Then:

$$\lambda \theta_T(\hat{t},\hat{q}_1) + (1-\lambda)\theta_T(\hat{t},\hat{q}_2) - \theta_T(\hat{t},\lambda \hat{q}_1 + (1-\lambda)\hat{q}_2)$$
$$\le \mathlarger{\int}_{\hat{t}}^{T} V_s \left(\lambda L\left(\frac{\dot{q}_1(s)}{V_s}\right) + (1-\lambda) L\left(\frac{\dot{q}_2(s)}{V_s}\right) - L\left(\frac{\dot{q}^*_{ \hat{t}, \lambda \hat{q}_1 + (1-\lambda)\hat{q}_2 }(s)}{V_s}\right) \right) ds$$
$$+ \frac 12 \gamma \sigma^2 \mathlarger{\int}_{\hat{t}}^{T}  \left(\lambda q_1(s)^2 + (1-\lambda) q_2(s)^2 - q^*_{ \hat{t}, \lambda \hat{q}_1 + (1-\lambda)\hat{q}_2 }(s)^2\right) ds.$$

Because $L$ is locally semi-concave, there exists a continuous and nondecreasing function $\omega$ with $\omega(0) = 0$ such that (for $\epsilon = T-\hat{t}$):\\

$\forall \rho_1,\rho_2 \in \left[-\frac{1}{\underline{V}}\left(C(Q,\epsilon) + 2\frac{Q}{\epsilon}\right), \frac{1}{\underline{V}}\left(C(Q,\epsilon) + 2\frac{Q}{\epsilon}\right)\right], \forall \lambda \in (0,1)$, $$\lambda L\left(\rho_1\right) + (1-\lambda) L\left(\rho_2\right) - L\left(\lambda \rho_1 + (1-\lambda)\rho_2 \right) \le \lambda(1-\lambda)|\rho_1-\rho_2|\omega(|\rho_1-\rho_2|).$$

Hence,

$$\lambda \theta_T(\hat{t},\hat{q}_1) + (1-\lambda)\theta_T(\hat{t},\hat{q}_2) - \theta_T(\hat{t},\lambda \hat{q}_1 + (1-\lambda)\hat{q}_2)$$$$ \le \lambda(1-\lambda) \mathlarger{\int}_{\hat{t}}^{T} V_s\left|\frac{\dot{q}_2(s)}{V_s}- \frac{\dot{q}_1(s)}{V_s}\right| \omega\left(\left|\frac{\dot{q}_2(s)}{V_s}- \frac{\dot{q}_1(s)}{V_s}\right|\right) ds$$
$$+ \frac 12 \gamma \sigma^2 \lambda (1-\lambda) (\hat{q}_2-\hat{q}_1)^2 \mathlarger{\int}_{\hat{t}}^{T} \left(\frac{T-s}{T-\hat{t}}\right)^2ds.$$

Consequently,

$$\lambda \theta_T(\hat{t},\hat{q}_1) + (1-\lambda)\theta_T(\hat{t},\hat{q}_2) - \theta_T(\hat{t},\lambda \hat{q}_1 + (1-\lambda)\hat{q}_2)$$$$ \le \lambda(1-\lambda) \left|\hat{q}_2-\hat{q}_1\right| \omega\left(\frac{\left|\hat{q}_2-\hat{q}_1\right|}{\epsilon \underline V} \right)+ \frac 16 \gamma \sigma^2 \lambda (1-\lambda) \epsilon (\hat{q}_2-\hat{q}_1)^2.$$

This proves the semi-concavity of $\theta_T(\hat{t},\cdot)$ on $[-Q,Q]$. This result being true for all $Q$, the local semi-concavity of $\theta_T(\hat{t},\cdot)$ is proved.\qed\\

Now, we prove the important result of this appendix stating that $\theta_T$ is in fact a $C^1$ function on $[0,T)\times\mathbb{R}$ solving the PDE of Proposition \ref{visco} in the classical sense.\\

\textbf{Theorem} (Regularity of $\theta_T$ and Hamilton-Jacobi equation)
Let us assume that $L$ is locally semi-concave.\\
Then, $\theta_T \in C^1([0,T)\times\mathbb{R})$ and:
$$\forall (t,q) \in[0,T) \times \mathbb{R},  -\partial_t \theta_T(t,q) - \frac 12 \gamma \sigma^2 q^2 + V_t H(\partial_q \theta_T(t,q)) = 0.$$

\emph{Proof:}\\

A first step in the proof consists of recalling that a function which is both locally semi-concave and convex is $C^1$. This gives $\forall t \in [0,T), \theta_T(t,\cdot) \in C^1(\mathbb R)$.\\
$\theta_T$ is $C^0$ on $[0,T) \times \mathbb{R}$ with $\forall t \in [0,T), \theta_T(t,\cdot)$ convex and $C^1$. Hence, we deduce from Theorem 25.7 of \cite{rockafellar1996convex} that $\partial_x \theta_T$ is in fact a continuous function on $[0,T) \times \mathbb{R}$.\\

Now, let us consider $q \in \mathbb{R}$. We know from Proposition \ref{lip} that $\theta_T(\cdot,q) \in W_{\mathrm{loc}}^{1,\infty}(0,T)$, and then it is almost everywhere differentiable on $(0,T)$. If $t$ is such that $\theta_T(\cdot,q)$ is differentiable at $t$, then $\theta_T$ is differentiable at $(t,q)$ and we know that $-\partial_t \theta_T(t,q) - \frac 12 \gamma \sigma^2 q^2 + V_t H(\partial_q \theta_T(t,q)) = 0$. But this gives that almost everywhere on $[0,T)$, $\partial_t \theta_T(\cdot,q)$ is equal to a continuous function. Therefore $\theta_T(\cdot,q)$ is in fact $C^1([0,T))$ with $\partial_t \theta_T(t,q) = - \frac 12 \gamma \sigma^2 q^2 + V_t H(\partial_q \theta_T(t,q))$. We conclude that $\partial_t \theta_T$ is a continuous function. Therefore $\theta_T \in C^1([0,T)\times\mathbb R)$ and $\theta_T$ solves the above PDE on $[0,T) \times \mathbb{R}$.\qed\\

\section*{Appendix B: Proofs}

\emph{Proof of Proposition \ref{dist}:}\\

By definition:
\begin{eqnarray*}
% \nonumber to remove numbering (before each equation)
  X_t &=& \int_0^t v_s S_s ds - \int_0^t V_s L\left(\frac{v_s}{V_s}\right)ds  - \psi \int_0^t |v_s| ds  \\
   &=& q_0 S_0 - q_tS_t - \int_0^t f(|q_0-q_s|) v_s q_s ds + \int_0^t \sigma q_s dW_s\\
    &&- \int_0^t V_s L\left(\frac{v_s}{V_s}\right) ds - \psi \int_0^t |v_s| ds  \\
   &=& q_0 S_0 -q_tS_t  - q_t \int_0^{q_0-q_t} f(|z|) dz - \int_0^{q_0-q_t} F(z) dz\\
    &&- \int_0^t V_s L\left(\frac{v_s}{V_s}\right) ds - \psi \int_0^t |v_s| ds + \int_0^t \sigma q_s dW_s.\\
\end{eqnarray*}
When $(q_s)_{s \in [0,t]}$ is deterministic, if $\int_0^t V_s L\left(\frac{v_s}{V_s}\right) ds < +\infty$ then  $X_t + q_t S_t$ is normally distributed with mean $q_0 S_0 - q_t \int_0^{q_0-q_t} f(|z|) dz - \int_0^{q_0-q_t} F(z) dz - \int_0^t V_s L\left(\frac{v_s}{V_s}\right) ds - \psi \int_0^t |v_s| ds$ and variance $\sigma^2 \int_0^t q^2_s ds$.\qed\\

\emph{Proof of Theorem \ref{minimizer}:}\\

Let us first consider the function $q_{line} \in AC_{q_0,0}(0,T)$ defined by $q_{line}(t) = q_0 \left(1- \frac tT\right)$. $\mathcal{I}_\psi\left(q_{line}\right) < +\infty$ and therefore $\inf_{AC_{q_0,0}(0,T)} \mathcal{I}_\psi < +\infty$. Also, given the hypotheses on $L$, it is clear that $\inf_{AC_{q_0,0}(0,T)} \mathcal{I}_\psi \ge 0$.\\

Now, let us consider a sequence $(q_n)_{n \in \mathbb{N}}$ of functions in $AC_{q_0,0}(0,T)$ such that $\lim_{n \to +\infty}\mathcal{I}_\psi(q_n) = \inf_{AC_{q_0,0}(0,T)} \mathcal{I}_\psi$.\\

The first step consists of showing that $(\dot{q}_n)_n$ is equiabsolutely integrable. For that purpose, we use the superlinear growth of the function $L$ and consider for $A > 0$, a constant $C_A$ such that $\forall y > C_A, \frac{L(y)}{A}>y$. Hence, if we consider a measurable set $E \subset [0,T]$, we have:
$$\int_E |\dot{q}_n(s)| ds \le \int_{E} \frac{ V_s L\left( \frac{|\dot{q}_n(s)|}{V_s}\right)}{A} 1_{\{|\dot{q}_n(s)|> C_A V_s\}}ds + C_A \overline{V} |E|.$$

Hence:

$$\int_E |\dot{q}_n(s)| ds \le \frac 1A \mathcal{I}_\psi(q_n)  + C_A \overline{V} |E|.$$

Since this can be made as small as possible by first choosing $A$ and then adjusting $|E|$, $(\dot{q}_n)_n$ is equiabsolutely integrable.\\

In particular, because $q_n(0) = q_0$, we know that, up to a subsequence, $(q_n)_n$ converges uniformly on $[0,T]$ towards a function $q^* \in AC_{q_0,0}(0,T)$ and that $\dot{q}_n \rightharpoonup \dot{q}^*$ (weak convergence in $L^1$) -- this is Dunford-Pettis theorem and Ascoli theorem.\\

Hence, we first have that

$$\frac 12 \gamma \sigma^2 \mathlarger{\int}_0^T  q_n(s)^2 ds \to_{n \to \infty} \frac 12 \gamma \sigma^2 \mathlarger{\int}_0^T  q^*(s)^2 ds.$$
Second, because of our convexity assumption on $L$, we know that
$$v \in L^1(0,T) \mapsto \left(\mathlarger{\int}_0^T V_s L\left(\frac{v(s)}{V_s}\right) ds + \psi \int_0^T |v(s)| ds\right)$$ is lower semi-continuous for the weak topology, and hence that:

$$\mathlarger{\int}_0^T V_s L\left(\frac{\dot{q}^{*}(s)}{V_s} \right) ds + \psi \int_0^T |\dot{q}^*(s)| ds \le \liminf_{n \to \infty}
\mathlarger{\int}_0^T V_s L\left(\frac{\dot{q}_n(s)}{V_s}\right) ds + \psi \int_0^T |\dot{q}_n(s)| ds.$$

Combining the inequalities, we get $\mathcal{I}_\psi(q^*) \le \liminf_{n \to \infty} \mathcal{I}_\psi(q_n)$, and therefore that $q^*$ is a minimizer.\\

Coming to uniqueness, if there were two different minimizers, $q^*$ and $q^{**}$, then we would have $\mathcal{I}_\psi\left(\frac{q^{*} + q^{**}}{2}\right) < \frac 12 \mathcal{I}_\psi(q^{*}) + \frac 12 \mathcal{I}_\psi(q^{**}) = \mathcal{I}_\psi(q^*)$ because $L$ is convex and $q \mapsto \frac 12 \gamma \sigma^2 q^2$ is strictly convex, in contradiction with the optimality of $q^*$. Hence, $q^*$ is the unique minimizer.\\

Now, we can define $\tilde{q}^*$ by $\tilde{q}^*(t) = \inf_{s \le t} (q^*(s))_+$. We have $\tilde{q}^* \in AC_{q_0,0}(0,T)$ with $\dot{\tilde{q}}^* = \dot{q}^* 1_{\lbrace q^* > 0, q^*=\tilde{q}^* \rbrace}$. Hence, using the monotonicity property of $L$, we obtain that $\mathcal{I}_\psi(\tilde{q}^*) \le \mathcal{I}_\psi(q^*)$.\\

Hence  $q^* = \tilde{q}^*$ and the minimizer is a nonnegative and nonincreasing function. In particular $q^*$ is the minimizer of:

$$q \mapsto \mathlarger{\int}_0^T \left(V_sL\left(\frac{\dot{q}(s)}{V_s}\right) + \frac 12 \gamma \sigma^2  q^2(s) \right) ds + \psi q_0 = \mathcal{I}_0(q) + \psi q_0.$$

Therefore, $q^*$ is the minimizer of $\mathcal{I}_0$ and it is independent of $\psi$.\qed\\

\emph{Proof of Theorem \ref{optisdet}:}\\

For any $v \in \mathcal{A}_0$, we know that $$X_T = q_0 S_0  - \int_0^{q_0} F(z) dz - \int_0^T V_s L\left(\frac{v_s}{V_s}\right) ds - \psi \int_0^T |v_s| ds + \int_0^T \sigma q_s dW_s.$$
Hence:

$$\mathbb{E}\left[-\exp\left(-\gamma X_T\right)\right] = -\exp\left(-\gamma\left(q_0 S_0  - \int_0^{q_0} F(z) dz\right)\right)$$ $$\times\mathbb{E}\left[\exp\left(\gamma \left(\int_0^T V_s L\left(\frac{v_s}{V_s}\right) ds + \psi \int_0^T |v_s| ds\right)\right)\exp\left(-\gamma \sigma\int_0^T q_s dW_s\right)\right].$$

Hence, if we introduce the probability measure $\mathbb Q$ defined by the Radon-Nikodym derivative\footnote{We can apply Girsanov theorem since $q$ is bounded.}
$$\frac{d\mathbb Q}{d\mathbb P} = \exp\left(-\gamma \sigma\int_0^T q_s dW_s - \frac 12 \gamma^2\sigma^2 \int_0^T q_s^2 ds\right),$$
then:

$$\mathbb{E}\left[-\exp\left(-\gamma X_T\right)\right] = -\exp\left(-\gamma\left(q_0 S_0  - \int_0^{q_0} F(z) dz\right)\right)$$$$ \times\mathbb{E}^{\mathbb Q}\left[\exp\left(\gamma \left(\int_0^T V_s L\left(\frac{v_s}{V_s}\right) ds + \psi \int_0^t |v_s| ds\right)\right)\exp\left(\frac 12 \gamma^2\sigma^2 \int_0^T q_s^2 ds\right)\right].$$

Hence: $$\mathbb{E}\left[-\exp\left(-\gamma X_T\right)\right] = \mathbb{E}^{\mathbb Q}\left[-\exp\left(-\gamma \left(q_0 S_0 - \int_0^{q_0} F(z) dz - \mathcal{I}_{\psi}(q)\right)\right)\right].$$

Now, almost surely over $\Omega$, $t \mapsto q_t(\omega)$ is absolutely continuous. Hence, almost surely, $\mathcal{I}_{\psi}(q(\omega)) \ge \mathcal{I}_{\psi}(q^*)$.\\
This leads to $$\mathbb{E}\left[-\exp\left(-\gamma X_T\right)\right] \le -\exp\left(-\gamma \left(q_0 S_0 - \int_0^{q_0} F(z) dz - \mathcal{I}_{\psi}(q^*)\right)\right),$$
\emph{i.e.}:
$$\mathbb{E}\left[-\exp\left(-\gamma X_T\right)\right] \le \sup_{v\in \mathcal{A}_{0,\mathrm{det}}} \mathbb{E}\left[-\exp\left(-\gamma X_T\right)\right].$$

We then obtain $$\sup_{v\in \mathcal{A}_0} \mathbb{E}\left[-\exp\left(-\gamma X_T\right)\right] \le \sup_{v\in \mathcal{A}_{0,\mathrm{det}}} \mathbb{E}\left[-\exp\left(-\gamma X_T\right)\right].$$

Since the converse inequality holds, the result is proved.\qed\\

In the following proofs, we denote $q^*_{\hat{t},\hat{q}} \in AC_{\hat{q},0}(\hat{t},T)$ the function at which the infimum (minimum) in the definition of $\theta_T(\hat{t},\hat{q})$ is attained. Such a function does exist by Theorem~\ref{minimizer} since we can replace the couple $(0,q_0)$ by the couple $(\hat{t},\hat{q})$ -- and it is a $C^1$ function on $[\hat{t},T]$.\\

\emph{Proof of Proposition \ref{mono}:}\\

Consider $t,\hat{t} \in [0,T)$ with $t<\hat{t}$. Let us define $q : s \in [t,T] \mapsto q^*_{\hat{t},\hat{q}}(s - t + \hat{t})1_{s \le T + t - \hat{t}}$.\\

We have:

\begin{eqnarray*}
% \nonumber to remove numbering (before each equation)
  \theta_T(t,\hat{q}) &\le& \int_t^{T} \left(V_sL\left(\frac{\dot{q}(s)}{V_s}\right)  + \frac 12 \gamma \sigma^2  q(s)^2 \right) ds \\
   &\le& \int_t^{T+t-\hat{t}} \left(V_sL\left(\frac{\dot{q}(s)}{V_s}\right)  + \frac 12 \gamma \sigma^2  q(s)^2 \right) ds  \\
   &\le&  \int_{\hat{t}}^{T} \left(V_sL\left(\frac{\dot{q}^*_{\hat{t},\hat{q}}(s)}{V_s}\right)  + \frac 12 \gamma \sigma^2  q^*_{\hat{t},\hat{q}}(s)^2 \right) ds  \\
   &\le & \theta_T(\hat{t},\hat{q}).\\
\end{eqnarray*}

Hence, $\theta_T(\cdot,\hat{q})$ is an nondecreasing function on $[0,T)$.\qed\\

\emph{Proof of Corollary \ref{como}:}\\

Consider $T,T' \in (t,+\infty)$ with $T<T'$. We have that $\theta_T(t,q) = \theta_{T'}(T'-T+t,q) \ge \theta_{T'}(t,q)$.\qed\\

\emph{Proof of Proposition \ref{monoq}:}\\

Consider $q,\hat{q} \in \mathbb{R}_+$ with $q < \hat{q}$. Then:

\begin{eqnarray*}
% \nonumber to remove numbering (before each equation)
  \theta_T(\hat{t},q) &\le& \int_t^{T} \left(V_sL\left(\frac{q}{\hat{q}}\frac{\dot{q}^*_{\hat{t},\hat{q}}(s)}{V_s}\right)  + \frac 12 \gamma \sigma^2  \left(\frac{q}{\hat{q}}q^*_{\hat{t},\hat{q}}(s)\right)^2 \right) ds \\
   &\le& \int_{\hat{t}}^{T} \left(V_sL\left(\frac{\dot{q}^*_{\hat{t},\hat{q}}(s)}{V_s}\right)  + \frac 12 \gamma \sigma^2  q^*_{\hat{t},\hat{q}}(s)^2 \right) ds  \\
   &\le & \theta_T(\hat{t},\hat{q}).\\
\end{eqnarray*}

Hence, $\theta_T(\hat{t},\cdot)$ is an nondecreasing function on $\mathbb{R}_+$.\qed\\

\emph{Proof of Proposition \ref{convex}:}\\

Let us consider $\hat{q} \in \mathbb{R}$ and $h>0$.

We consider $q = \frac 12 q^*_{ \hat{t},\hat{q}-h } + \frac 12 q^*_{ \hat{t},\hat{q}+h }$ and we have:

$$\theta_T(\hat{t},\hat{q} + h) - 2\theta_T(\hat{t},\hat{q}) + \theta_T(\hat{t},\hat{q} - h)$$$$ \ge \mathlarger{\int}_{\hat{t}}^{T} V_s \left(L\left(\frac{\dot{q}^*_{ \hat{t},\hat{q} +h }(s)}{V_s}\right) - 2L\left(\frac{\dot{q}(s)}{V_s}\right) + L\left(\frac{\dot{q}^*_{ \hat{t},\hat{q} -h }(s)}{V_s}\right) \right) ds$$
$$+ \frac 12 \gamma \sigma^2 \mathlarger{\int}_{\hat{t}}^{T}  \left(q^*_{ \hat{t},\hat{q}+h }(s)^2 - 2 q(s)^2 + q^*_{ \hat{t},\hat{q}-h}(s)^2\right) ds.$$

Using the convexity of $L$ and the convexity of $x \to x^2$, we obtain that:

$$\theta_T(\hat{t},\hat{q} + h) - 2\theta_T(\hat{t},\hat{q}) + \theta_T(\hat{t},\hat{q} - h) \ge 0.$$

Hence, $\theta_T(\hat{t},\cdot)$ is a convex function. \qed\\

\emph{Proof of Proposition \ref{lip}:}\\

In order to prove this Proposition, we need the following Lemma that provides a bound to the gradient of $q^*_{\hat{t},\hat{q}}$:\\

\textbf{Lemma}\\
$$\forall Q > 0, \forall \epsilon \in (0,T), \exists C(Q,\epsilon) >0, \forall (\hat{t},\hat{q}) \in [0,T-\epsilon]\times [-Q,Q], \sup_{t \in [\hat{t},T]} |\dot{q}^*_{\hat{t},\hat{q}}(t)| \le C(Q,\epsilon).$$

\emph{Proof of the Lemma:}\\

We consider the case $\hat{q} \ge 0$. The case of a negative $\hat{q}$ works exactly the same.\\

Using the same method as for Proposition \ref{hamilton}, we know that $q^*_{\hat{t},\hat{q}}$ solves an ODE of the form:

\[
\left \{
\begin{array}{c c c}
    \dot{p}(t) & = & \gamma \sigma^2 q(t) \\
    \dot{q}(t) & = & V_t H'(p(t))\\
\end{array}
\right. \qquad  q(\hat{t}) = \hat{q}, \quad q(T) = 0
\]
where $p$ is a $C^1$ function.

From the ODE, we know that $\forall t \in [\hat{t},T], p(t) \in [p(\hat{t}), p(\hat{t}) + \gamma \sigma^2 (T - \hat{t})\hat{q}]$. This gives $\forall t \in [\hat{t},T], \dot{q}^*_{\hat{t},\hat{q}}(t) \in \left[\overline{V}H'(p(\hat{t})), \min\left(0,\underline{V}H'(p(\hat{t}) + \gamma \sigma^2 (T - \hat{t})\hat{q})\right)\right]$.\\

Now, if $H'(p(\hat{t}) + \gamma \sigma^2 (T - \hat{t}) \hat{q}) \ge 0$, then $H'(p(\hat{t}) + \gamma \sigma^2 QT) \ge 0$ and $p(\hat{t})$ is bounded from below by a constant that depends on $Q$.\\

Otherwise, we have:

$$- \hat{q} = \int_{\hat{t}}^{T} \dot{q}^*_{\hat{t},\hat{q}}(t) dt \le \underline{V}H'(p(\hat{t}) + \gamma \sigma^2 (T - \hat{t})\hat{q}) (T - \hat{t}) \le \underline{V}H'(p(\hat{t}) + \gamma \sigma^2 Q T) (T - \hat{t}).$$

Hence $H'(p(\hat{t}) + \gamma \sigma^2 Q T) \ge -\frac{\hat{q}}{\underline{V}(T-\hat{t})} \ge -\frac Q{\epsilon\underline{V}}$ and $p(\hat{t})$ is bounded from below by a constant that depends on $Q$ and $\epsilon$.\\

Now, in general, because $\forall t \in [\hat{t},T], \dot{q}^*_{\hat{t},\hat{q}}(t) \ge \overline{V}H'(p(\hat{t}))$, we have a lower bound for $\dot{q}^*_{\hat{t},\hat{q}}(t)$. $0$ being an upper bound, we have proved the result.\qed\\

Given $Q > 0$ and  $\epsilon \in (0,T)$, this lemma and the results of Section 2 allow to redefine $\theta_T(\hat{t},\hat{q})$ for $(\hat{t},\hat{q}) \in [0,T-\epsilon]\times [-Q,Q]$ by:$$\theta_T(\hat{t},\hat{q}) = \inf_{q \in C^1([\hat{t},T]), q(\hat{t}) = \hat{q}, q(T) = 0, |\dot{q}(t)| \le C(Q,\epsilon)} \mathlarger{\int}_{\hat{t}}^T \left(V_sL\left(\frac{\dot{q}(s)}{V_s}\right)  + \frac 12 \gamma \sigma^2  q^2(s) \right) ds.$$

\emph{Proof the Proposition:}\\

Let us consider $(\hat{t},\hat{q})$ and $(\tilde{t},\tilde{q})$ in $[0,T-\epsilon]\times[-Q,Q]$.\\

From the convexity of $\theta_T(\tilde{t},\cdot)$, we know that there exists $K(Q,\epsilon)$ such that $\forall (q,\overline{q}) \in [-Q,Q]^2, |\theta_T(\tilde{t},q) - \theta_T(\tilde{t},\overline{q})| \le K(Q,\epsilon) |q - \overline{q}|$.\\

Now, if $\hat{t} < \tilde{t}$:

$$\theta_T(\hat{t},\hat{q}) = \mathlarger{\int}_{\hat{t}}^{\tilde{t}} \left(V_sL\left(\frac{\dot{q}^*_{\hat{t},\hat{q}}(s)}{V_s}\right)  + \frac 12 \gamma \sigma^2  q^*_{\hat{t},\hat{q}}(s)^2 \right) ds + \theta_T(\tilde{t},q^*_{\hat{t},\hat{q}}(\tilde{t})).$$

Since $\dot{q}^*_{\hat{t},\hat{q}}$ and $q^*_{\hat{t},\hat{q}}$ are uniformly bounded, there exists a constant $M(Q,\epsilon)$ such that:

$$\theta_T(\hat{t},\hat{q}) \le M(Q,\epsilon) (\tilde{t} - \hat{t}) + \theta_T(\tilde{t},q^*_{\hat{t},\hat{q}}(\tilde{t})) \le M(Q,\epsilon) (\tilde{t} - \hat{t}) + \theta_T(\tilde{t},\hat{q}) + K(Q,\epsilon) |q^*_{\hat{t},\hat{q}}(\tilde{t})-\hat{q}|$$$$ \le M(Q,\epsilon) (\tilde{t} - \hat{t}) + \theta_T(\tilde{t},\hat{q}) + K(Q,\epsilon) C(Q,\epsilon)|\tilde{t}-\hat{t}|.$$

Hence, there exists a constant $K'(Q,\epsilon)$ such that:

$$0 \le \theta_T(\hat{t},\hat{q}) - \theta_T(\tilde{t},\hat{q}) \le K'(Q,\epsilon) |\tilde{t}-\hat{t}|.$$

Combining the results we eventually obtain:

$$|\theta_T(\hat{t},\hat{q}) - \theta_T(\tilde{t},\tilde{q})| \le K(Q,\epsilon) |\tilde{q}-\hat{q}| + K'(Q,\epsilon) |\tilde{t}-\hat{t}|.$$\qed\\

\emph{Proof of Proposition \ref{visco}:}\\

We start with the fact that $\theta_T$ is a subsolution of the Hamilton-Jacobi equation on $(0,T)\times\mathbb{R}$.\\

Let $(t,q) \in (0,T)\times\mathbb{R}$ and let $\varphi \in C^1((0,T)\times\mathbb{R})$ be such that $\theta_T - \varphi$ has a local maximum in $(t,q)$. We can then find $\tau>0$ and $\eta>0$ such that $\forall (t',q') \in [t-\tau,t+\tau]\times[q-\eta,q+\eta]:$
$$\theta_T(t,q) - \varphi(t,q) \ge \theta_T(t',q') - \varphi(t',q').$$

Now, let us consider $v \in \mathbb{R}$. If $h \in (0,\tau]$ is small enough, we have that $t+h < T$ and $q-vh \in [q-\eta,q+\eta]$. Then:

\begin{eqnarray*}
% \nonumber to remove numbering (before each equation)
  \varphi(t+h,q-vh) - \varphi(t,q) &\ge& \theta_T(t+h,q-vh) - \theta_T(t,q) \\
  \frac{\varphi(t+h,q-vh) - \varphi(t,q)}{h} &\ge& -\frac{1}{h} \int_t^{t+h} \left(V_sL\left(\frac{v}{V_s}\right)  + \frac 12 \gamma \sigma^2  (q-v(s-t))^2 \right) ds.  \\
\end{eqnarray*}
Hence if $h \to 0$, we get:

$$\partial_t \varphi(t,q) - v \partial_q \varphi(t,q) \ge -V_tL\left(\frac{v}{V_t}\right)  - \frac 12 \gamma \sigma^2  q^2.$$

Hence, $\forall v\in \mathbb{R}$,

$$ -\partial_t \varphi(t,q) - \frac 12 \gamma \sigma^2  q^2 + v \partial_q \varphi(t,q)  - V_tL\left(\frac{v}{V_t}\right) \le 0.$$
This gives:
$$ -\partial_t \varphi(t,q) - \frac 12 \gamma \sigma^2  q^2 + \sup_{v \in \mathbb{R}} v \partial_q \varphi(t,q)  - V_tL\left(\frac{v}{V_t}\right) \le 0$$
$$ -\partial_t \varphi(t,q) - \frac 12 \gamma \sigma^2  q^2 + V_t\sup_{v \in \mathbb{R}} \frac{v}{V_t} \partial_q \varphi(t,q)  - L\left(\frac{v}{V_t}\right) \le 0$$
$$ -\partial_t \varphi(t,q) - \frac 12 \gamma \sigma^2  q^2 + V_tH\left(\partial_q \varphi(t,q)\right)\le 0.$$

This proves that $\theta_T$ is a subsolution of the Hamilton-Jacobi equation on $(0,T)\times \mathbb{R}$.\\

We now prove that $\theta_T$ is a supersolution of the Hamilton-Jacobi equation on $(0,T)\times\mathbb{R}$.\\

Let $(t,q) \in (0,T)\times\mathbb{R}$ and let $\varphi \in C^1((0,T)\times\mathbb{R})$ be such that $\theta_T - \varphi$ has a local minimum in $(t,q)$. We can then find $\tau>0$ and $\eta>0$ such that $\forall (t',q') \in [t-\tau,t+\tau]\times[q-\eta,q+\eta]:$
$$\theta_T(t,q) - \varphi(t,q) \le \theta_T(t',q') - \varphi(t',q').$$

$\forall \epsilon>0, \forall h \in (0,\min(\tau,\frac{T-t}{2})), \exists (v^{\epsilon,h}_s)_{s \in [t,T]} \in \mathcal{A}_{0,\mathrm{det}}(t,q)$ such that:

$$\theta_T(t,q) - \theta_T\left(t+h,q^{v^{\epsilon,h}}_{t+h}\right) \ge \int_t^{t+h} \left(V_sL\left(\frac{v^{\epsilon,h}_s}{V_s}\right)  + \frac 12 \gamma \sigma^2  (q^{v^{\epsilon,h}}_s)^2 \right) ds  - \epsilon h,$$
where $q^{v^{\epsilon,h}}_s = q - \int_t^s v^{\epsilon,h}_w dw$.\\

Now, because of the above Lemma, we can always suppose that $|v^{\epsilon,h}|$ is uniformly bounded, independently of $\epsilon$ and $h$, by a constant $M$. Hence, if we have $h < \tau$, $h < \frac{T-t}{2}$, and $Mh < \eta$, then $q^{v^{\epsilon,h}}_{t+h} \in  [q-\eta,q+\eta]$ and:

$$ \varphi(t,q) - \varphi\left(t+h,q^{v^{\epsilon,h}}_{t+h}\right) \ge \int_t^{t+h} \left(V_sL\left(\frac{v^{\epsilon,h}_s}{V_s}\right)  + \frac 12 \gamma \sigma^2  (q^{v^{\epsilon,h}}_s)^2 \right) ds - \epsilon h.$$
This gives:
$$-\int_t^{t+h} \left(\partial_t \varphi\left(s,q^{v^{\epsilon,h}}_{s}\right) - v^{\epsilon,h}_{s} \partial_q \varphi\left(s,q^{v^{\epsilon,h}}_{s}\right) \right) ds$$$$ \ge \int_t^{t+h} \left(V_sL\left(\frac{v^{\epsilon,h}_s}{V_s}\right)  + \frac 12 \gamma \sigma^2  (q^{v^{\epsilon,h}}_s)^2 \right) ds - \epsilon h.$$
Hence:
$$\int_t^{t+h} \left(-\partial_t \varphi\left(s,q^{v^{\epsilon,h}}_{s}\right) - \frac 12 \gamma \sigma^2  (q^{v^{\epsilon,h}}_s)^2 + v^{\epsilon,h}_{s} \partial_q \varphi\left(s,q^{v^{\epsilon,h}}_{s}\right) - V_sL\left(\frac{v^{\epsilon,h}_s}{V_s}\right) \right) ds \ge - \epsilon h$$
$$\int_t^{t+h} \left(-\partial_t \varphi\left(s,q^{v^{\epsilon,h}}_{s}\right) - \frac 12 \gamma \sigma^2  (q^{v^{\epsilon,h}}_s)^2  + V_sH\left(\partial_q \varphi\left(s,q^{v^{\epsilon,h}}_{s}\right)\right) \right) ds \ge - \epsilon h$$
$$\frac 1h\int_t^{t+h} \left(-\partial_t \varphi\left(s,q^{v^{\epsilon,h}}_{s}\right) - \frac 12 \gamma \sigma^2  (q^{v^{\epsilon,h}}_s)^2 + V_sH\left(\partial_q \varphi\left(s,q^{v^{\epsilon,h}}_{s}\right)\right) \right) ds \ge - \epsilon. $$

Using, the continuity of $\partial_t \varphi$, $\partial_q \varphi$, and $H$, and the uniform bound on $v^{\epsilon,h}$, we then obtain, sending $h$ to $0$, that:

$$-\partial_t \varphi\left(t,q\right) - \frac 12 \gamma \sigma^2  q^2 + V_tH\left(\partial_q \varphi\left(t,q\right)\right) \ge - \epsilon.$$

This being true for any $\epsilon >0$, we have that:
$$-\partial_t \varphi\left(t,q\right) - \frac 12 \gamma \sigma^2  q^2 + V_tH\left(\partial_q \varphi\left(t,q\right)\right) \ge 0$$

This proves that $\theta_T$ is a supersolution of the Hamilton-Jacobi equation on $(0,T)\times \mathbb{R}$.\\

Consequently, $\theta_T$ is a viscosity solution of the Hamilton-Jacobi equation on $(0,T)\times \mathbb{R}$.

We need to prove now that the result holds on $[0,T)\times \mathbb{R}$.\\

Let $q \in \mathbb{R}$ and let $\varphi \in C^1([0,T)\times \mathbb{R})$ be such that $\theta_T - \varphi$ has a strict local maximum in $(0,q)$. Then, for $\epsilon > 0$ sufficiently small, $(t,q) \in (0,T)\times \mathbb{R} \mapsto \theta_T(t,q) - \varphi(t,q) - \frac{\epsilon}{t}$ has a local maximum in $(t_\epsilon,q_{\epsilon}) \in (0,T)\times \mathbb{R}$, and $\lim_{\epsilon \to 0} (t_\epsilon,q_{\epsilon}) = (0,q)$. Using what we already proved, we obtain:

$$-\partial_t \varphi(t_\epsilon,q_\epsilon) + \frac{\epsilon}{t_\epsilon^2} - \frac 12 \gamma \sigma^2  q_\epsilon^2 + V_{t_\epsilon}H\left(\partial_q \varphi(t_\epsilon,q_\epsilon)\right)\le 0.$$
Hence:
$$-\partial_t \varphi(t_\epsilon,q_\epsilon) - \frac 12 \gamma \sigma^2  q_\epsilon^2 + V_{t_\epsilon}H\left(\partial_q \varphi(t_\epsilon,q_\epsilon)\right)\le 0,$$ and sending $\epsilon$ to $0$ we obtain:

$$-\partial_t \varphi(0,q) - \frac 12 \gamma \sigma^2  q^2 + V_0H\left(\partial_q \varphi(0,q)\right)\le 0.$$

Therefore, $\theta_T$ is a subsolution of the Hamilton-Jacobi equation on $[0,T)\times \mathbb{R}$.\\

Similarly, let $q \in \mathbb{R}$ and let $\varphi \in C^1([0,T)\times \mathbb{R})$ be such that $\theta_T - \varphi$ has a strict local minimum in $(0,q)$. Then, for $\epsilon > 0$ sufficiently small, $(t,q) \in (0,T)\times \mathbb{R} \mapsto \theta_T(t,q) - \varphi(t,q) + \frac{\epsilon}{t}$ has a local minimum in $(t_\epsilon,q_{\epsilon}) \in (0,T)\times \mathbb{R}$, and $\lim_{\epsilon \to 0} (t_\epsilon,q_{\epsilon}) = (0,q)$. Using what we already proved we obtain:

$$-\partial_t \varphi(t_\epsilon,q_\epsilon) - \frac{\epsilon}{t_\epsilon^2} - \frac 12 \gamma \sigma^2  q_\epsilon^2 + V_{t_\epsilon}H\left(\partial_q \varphi(t_\epsilon,q_\epsilon)\right)\ge 0.$$
Hence:
$$-\partial_t \varphi(t_\epsilon,q_\epsilon) - \frac 12 \gamma \sigma^2  q_\epsilon^2 + V_{t_\epsilon}H\left(\partial_q \varphi(t_\epsilon,q_\epsilon)\right)\ge 0,$$ and sending $\epsilon$ to $0$ we obtain:

$$-\partial_t \varphi(0,q) - \frac 12 \gamma \sigma^2  q^2 + V_0H\left(\partial_q \varphi(0,q)\right)\ge 0.$$

Therefore, $\theta_T$ is a supersolution of the Hamilton-Jacobi equation on $[0,T)\times \mathbb{R}$.\qed\\

\emph{Proof of Proposition \ref{singularity}:}\\

$\theta_T(\hat{t},0) = 0$. The result is then obvious for $\hat{q}=0$.\\

Now, using the properties of $L$, we get:
\begin{eqnarray*}
% \nonumber to remove numbering (before each equation)
  \theta_T(\hat{t},\hat{q}) &\ge & \mathlarger{\int}_{\hat{t}}^T V_sL\left(\frac{\dot{q}^*_{\hat{t},\hat{q}}(s)}{V_s}\right) ds \\
   &\ge& \mathlarger{\int}_{\hat{t}}^T \underline{V} L\left(\frac{\dot{q}^*_{\hat{t},\hat{q}}(s)}{\overline{V}}\right) ds \\
   &\ge & (T-\hat{t})\underline{V} L\left(\mathlarger{\int}_{\hat{t}}^T \frac{\dot{q}^*_{\hat{t},\hat{q}}(s)}{\overline{V}} \frac{ds}{T-\hat{t}}\right) \\
   &\ge & \underline{V}(T-\hat{t}) L\left(\frac{\hat{q}}{\overline{V}(T-\hat{t})}\right)
\end{eqnarray*}

Hence, if $\hat{q} \not= 0$, the superlinearity of $L$ gives $\lim_{\hat{t} \to T} \theta_T(\hat{t},\hat{q}) = +\infty$.\qed\\

\emph{Proof or Theorem \ref{asympt}:}\\

For $(t,q) \in \mathbb{R}_+ \times \mathbb{R}_+$, $T > t \mapsto \theta_T(t,q)$ is a nonincreasing function bounded from below by $0$. Hence $\lim_{T \to +\infty} \theta_T(t,q)$ exists.\\

Since for $0\le t' < t$, $\theta_T(t',q) = \theta_{T-t'+t}(t,q)$, $\lim_{T \to +\infty} \theta_T(t,q)$ is in fact independent of $t$ and we define $\theta_\infty(q) = \lim_{T \to +\infty} \theta_T(t,q)$.\\

Using Proposition~\ref{convex}, $\theta_\infty$ is a convex function and it is therefore continuous.\\

Hence, for a fixed $t \ge 0$, using Dini's theorem, the convergence of $(\theta_T(t,\cdot))_T$ towards $\theta_\infty$ is locally uniform on $\mathbb{R}_+$. Now, because of Proposition \ref{mono}, we have that $(\theta_T)_T$ converges towards $\theta_\infty$ locally uniformly on $\mathbb{R}_+\times\mathbb{R}_+$.\\

Now, we use the Hamilton-Jacobi equation of Proposition~\ref{visco} to prove that $\theta_\infty$ is a viscosity solution of:

$$- \frac 12 \gamma \sigma^2 q^2 + V H(\theta_\infty'(q)) = 0, \quad q > 0 , \quad \theta_\infty(0) = 0 \qquad (*).$$

We consider $q > 0$ and $\varphi \in C^1(\mathbb{R}_+^*)$ such that $\theta_\infty - \varphi$ has a local strict maximum in $q$. Then, we consider a sequence of triplets $(T_n,t_n,q_n) \in \mathbb{R}_+^2\times\mathbb{R}_+^*$ with $t_n < T_n$ such that $\lim_n T_n = +\infty$, $(t_n,q_n)_n$ converges with $\lim_n q_n = q$, and $\theta_{T_n} - \varphi$ has a local maximum in $(t_n,q_n)$. We therefore obtain that:
$$- \frac 12 \gamma \sigma^2 q_n^2 + V H(\varphi'(q_n)) \le 0.$$
Hence, considering the limit $n\to +\infty$, we have:
$$- \frac 12 \gamma \sigma^2 q^2 + V H(\varphi'(q)) \le 0.$$

Conversely, if we consider $q > 0$ and $\varphi \in C^1(\mathbb{R}_+^*)$ such that $\theta_\infty - \varphi$ has a local minimum in $q$. Then, we consider a sequence of triplets $(T_n,t_n,q_n) \in \mathbb{R}_+^2\times\mathbb{R}_+^*$ with $t_n < T_n$ such that $\lim_n T_n = +\infty$, $(t_n,q_n)_n$ converges with $\lim_n q_n = q$, and $\theta_{T_n} - \varphi$ has a local minimum in $(t_n,q_n)$. We therefore obtain that:
$$- \frac 12 \gamma \sigma^2 q_n^2 + V H(\varphi'(q_n)) \ge 0.$$
Hence, considering the limit $n\to +\infty$, we have:
$$- \frac 12 \gamma \sigma^2 q^2 + V H(\varphi'(q)) \ge 0.$$

This, and the fact that $\theta_\infty(0) = 0$, proves that $\theta_\infty$ is indeed a viscosity solution of the equation $(*)$.\\

Now, $\theta_\infty$ is convex and hence we have almost everywhere that $- \frac 12 \gamma \sigma^2 q^2 + V H(\theta_\infty'(q)) = 0$. Since $\theta_\infty$ is an nondecreasing function (as a limit of such functions), we can write that almost everywhere:

$$ \theta_\infty'(q) = H^{-1}\left(\frac 1{2V} \gamma \sigma^2 q^2 \right),$$ where $H^{-1}$ is the inverse of $H : \mathbb{R}_+ \to \mathbb{R}_+$, as defined in the above statement.\\

Since $\theta_\infty$ is convex we can integrate this equation to obtain:

$$\theta_\infty(q) = \theta_\infty(0) + \int_0^{q} H^{-1}\left(\frac{\gamma \sigma^2}{2V} x^2\right) dx = \int_0^{q} H^{-1}\left(\frac{\gamma \sigma^2}{2V} x^2\right) dx,$$ and the result is proved.\qed\\

\emph{Proof of Proposition \ref{sss}:}\\

When $L(\rho) = \eta |\rho|^{1+\phi}$, the function $H$ is given by $H(p) = \frac{\phi}{(1+\phi)^{1+\frac 1\phi}} \frac 1{\eta^{\frac 1\phi}} |p|^{1+\frac 1\phi}$. Hence, for $x\ge 0$, $H^{-1}(x) =\eta^{\frac 1{1+\phi}} \frac{1+\phi}{\phi^{\frac{\phi}{1+\phi}}} x^{\frac{\phi}{1+\phi}}$.\\ Using now the result of Theorem~\ref{closed}, we obtain the formula.\qed\\

\bibliographystyle{plain}

\end{document}